\documentclass[aps,twocolumn,showpacs,amsmath,amssymb,prd]{revtex4-1}
\usepackage{color}
\usepackage{graphicx}
\usepackage{bm}
\usepackage{subfigure}
\usepackage{longtable}
\makeatother

\begin{document}

\title{On stability of the Kasner solution in quadratic gravity}

\author{A. Toporensky}
\email{atopor@rambler.ru}
\affiliation{
Sternberg Astronomical Institute, Moscow Universiry, Moscow 119991, Russia\\
Kazan Federal University, Kremlevskaya 18, Kazan 420008, Russia
}
\author{D. M\"uller }

\email{muller@fis.unb.br}

\affiliation{Instituto de F\'{i}sica, Universidade de Bras\'{i}lia, Caixa Postal
04455, 70919-970 Bras\'{i}lia, Brazil}
\begin{abstract}
We consider dynamics of a flat anisotropic Universe 
filled by a perfect fluid near a cosmological
singularity in quadratic gravity. Two possible regimes are described --
the Kasner anisotropic solution and an isotropic ``vacuum radiation" solution
which has three sub cases depending on whether the equation of state parameter
$w$ is bigger, smaller or equals to $1/3$. Initial conditions for numerical integrations
have been chosen near General Relativity anisotropic solution with matter
(Jacobs solution). We have found that for such initial conditions there is a range
of values of coupling constants so that the resulting cosmological singularity
is isotropic.
\end{abstract}


\maketitle
\global\long\def\imsize{0.83\columnwidth}
 \global\long\def\halfsize{0.45\columnwidth}
\global\long\def\big{1.2\columnwidth}
\global\long\def\peq{0.65\columnwidth}

\section*{Introduction}
Studies of quadratic gravity is an area of intense investigations during last several decades. Initial motivation was to match gravity
with quantum issues on  semiclassical approach in a direct analogy with electromagnetism.

Today it's well known that for sufficiently strong fields, electromagnetism, is accordingly modified, due to vacuum polarization of the fermion field. The electromagnetic field is considered as a classical vector field, and  the fermion field is second quantized. This electromagnetism was initially found by Heisenberg and Euler \cite{Heisenberg:1935qt}, and also obtained by the method of Schwinger \cite{Schwinger:1951nm}. The result is a non linear electromagnetic theory.

Subsequently it was developed by De Witt \cite{Utiyama:1962sn,dewitt1967dynamical} in the context of a classically curved space time and a quantized field. The reason that it is known as the Schwinger - De Witt technique, see for example \cite{Birrell:1982ix,grib1994vacuum}. 

Much in the same way as in the electromagnetic theory, GR is modified to a quadratic in curvature theory
\begin{equation}
\mathcal{L}=\sqrt{-g}\left[\frac{1}{G}\left(R-\Lambda\right)+\alpha\left(R_{ab}R^{ab}-\frac{1}{3}R^{2}\right)+\beta R^{2}\right],\label{acao}
\end{equation}
which we will refer to as quadratic gravity (QG). 
A theory without these counter terms is inconsistent from the point of view of renormalization. Also the quadratic terms in the curvature improve the degree of the superficial divergence of the theory. Anyway, as is well known, the linearized degrees of freedom show a massless spin 2 field, a massive spin 0 field and a massive spin 2 field that has an energy with the wrong sign, see for example \cite{Stelle:1977ry, vanDam:1970vg}. The stability of the vacuum can be recovered at the expense of unitarity loss \cite{Stelle:1976gc}, which poses a major problem on the quantized version of this gravitational theory. 

On the other hand, for a strictly classical theory, the presence of a ghost is not so unacceptable. It only means that there will be non linear instabilities, which we do know that occur very often in nature. In this work, only the classical version of the theory will be addressed, the spirit is that it could well be an effective theory valid for some energy range. 

Quadratic gravity was first investigated by H. Weyl in 1918 \cite{weyl1918gravitation}, then it appeared again in the 60's \cite{buchdahl1962gravitational} through Buchdahl's seminal paper, followed by \cite{ruzmaikina1970quadratic}. It was popularized by Starobinsky  in connection to inflation \cite{gurovich1979quantum,starobinsky1980new}. For a good historical review see \cite{schmidt2007fourth}. 

Since then it has been investigated by many authors 
\cite{tomita1978anisotropic,Muller:1987hp,Berkin:1991nb,Barrow:2005qv,Barrow:2006xb,vitenti2006numerical,muller2006starobinsky,cotsakis2008slice,Barrow:2009gx,muller2011homogeneous,de2012bianchi,Muller:2012wx,MULLER:2014jaa,0264-9381-27-22-225013,PhysRevD.77.103523,PhysRevD.75.123515}. 
It is worth to note that cubic and higher order curvature corrections often lead to Big Rip singularity even in the framework of $f(R)$ theory
\cite {Carloni:2007br, Amendola:2006we, Ivanov:2011np, Bukzhalev:2013saa} or at least make important cosmological solutions unstable (for an example beyond the $f(R)$ theory see,
for example \cite{Sami:2005zc}), so our restriction with only quadratic curvature corrections is justified from phenomenological reasons also. In the context of Kasner like solutions in quadratic gravity, we must mention \cite{0264-9381-23-1-L01,0264-9381-23-9-011}. 


In this work we are going to focus on the Kasner  \cite{Kasner:1921zz}
solution which is an exact solution for quadratic gravity. The interest in this solution is that 
despite it is a vacuum solution, in GR it represent a past attractor for a Bianchi I Universe filled by a perfect fluid unless the stiff fluid
case (near a cosmological singularity ``matter does not matter"). Moreover, Kasner solutions are ``building blocks" for BKL chaotic approach
to a singularity in more general models like Bianchi IX mode \cite{Belinsky:1970ew}l. The study of asymptotic solutions in quadratic gravity was previously addressed, for example, by  \cite{Cotsakis:2007un, Cotsakis:1997ck, Miritzis:2003eu, Miritzis:2007yn}.

It was already known that the Kasner solution has a zero eigenvalue \cite{Barrow:2006xb} for quadratic gravity. The intention of this work is to understand what happens with the solution space near this zero eigenvector, and specifically at the space time singularity. We assume that
the Universe is filled by a perfect fluid and study how it approaches a cosmological singularity. As the Kasner solution is not a unique vacuum
solution in quadratic gravity (see bellow) the dynamics is not trivial.

The article is organized as follows. In section \ref{s1} a quick review of expansion normalized variables is given. This set of variables is used to write down the field equations  in \ref{s2}. In \ref{s3} we discuss past attractors of this system. In \ref{s4} our initial conditions are specified.
The section \ref{s5} contains a detailed description of numerical results obtained. The conclusions summarize our findings. 
For numerical codes we used gnu/gsl ode package, explicit embedded Runge-Kutta Prince-Dormand (8, 9) on linux. The codes were obtained using the algebraic manipulator Maple 16. In this work we choose the velocity of light $c=1$, while keep the coupling constant $G$ arbitrary. 
\section{Expansion Normalized Variables\label{s1}}

First of all, instead of the proper time  $s$,
$ds=dt/H(t)$, the dynamics is with respect to the dynamical time $t$, which is dimensionless. This choice of lapse function, and the meaning
of $H(t)$ will be made clear below.

An orthogonal non rigid base is used to define the variables  
\begin{eqnarray*}
 &  & ds^{2}=-\frac{dt^{2}}{H(t)^{2}}+\delta_{ij}\omega^{i}\otimes\omega^{j}\\
 &  & \delta_{ij}=e_{i}^{\mu}e_{j}^{\nu}g_{\mu\nu},
\end{eqnarray*}
where the $i,\, j$ indices refer to the spatial part and $\omega^{j}$
is the dual of $e_{i}$, $\omega^{j}e_{i}=\delta_{\, i}^{j}$. The
metric $g_{ab}$ is collectively called 
\begin{equation}
g_{ab}=\left(\begin{array}{cccc}
-H(t)^{-2} & 0 & 0 & 0\\
0 & 1 & 0 & 0\\
0 & 0 & 1 & 0\\
0 & 0 & 0 & 1
\end{array}\right).\label{metrica}
\end{equation}

For a timelike vector $u^{a}=(H,0,0,0),\; u^{a}u_{a}=-1$, the projection of the metric on
the 3-space orthogonal to $u^{a}$ is  
\[
h_{ab}=u_{a}u_{b}+g_{ab},\;\; h_{\, b}^{c}g_{ca}=h_{ab},\;\; h_{\, a}^{c}u_{c}=0,
\]
as is very well known from standard textbooks. The covariant derivative can be written in its irreducible parts
\begin{eqnarray}
 &  & \nabla_{a}u_{b}=\sigma_{ab}+\omega_{ab}+\frac{1}{3}\theta\delta_{ab}-\dot{u}_{a}u_{b}\nonumber \\
 &  & \sigma_{ab}=u_{(a;b)}-\frac{1}{3}\theta\delta_{ab}+\dot{u}_{(a}u_{b)}\nonumber \\
 &  & \omega_{ab}=u_{[a;b]}+\dot{u}_{[a}u_{b]}\nonumber \\
 &  & \dot{u_{a}}=u^{b}\nabla_{b}u_{a}\nonumber \\
 &  & \theta=\nabla_{c}u^{c},\label{teta}
\end{eqnarray}
where as promised, the lapse function given in \eqref{metrica} is
related to the expansion $\theta$ as 
\begin{equation}
H=\frac{\theta}{3}.\label{H}
\end{equation}
For spatially homogeneous spacetimes,
the timelike vector is geodesic $\dot{u}^{a}=0$ with zero vorticity
$\omega_{ab}=0$, being normal to the time slices. The temporal part of the connection and  the shear are related as 
\begin{eqnarray}
& &\nabla_{a}u_{b}  =  \delta_{a0}\delta_{b0}\frac{\dot{H}}{H}-\Gamma_{\, ab}^{0}u_{0}\label{conexao_0}\\
 &  & \Gamma_{\, ab}^{0}=\left\{ \begin{array}{l}
0,\, \mbox{if }a=0\;\mbox{or}/\mbox{and }b=0\\
H^{2}\sigma_{ij}+H^{2}\delta_{ij},\;\mbox{if }a\neq0\;\mbox{and }b\ne0
\end{array}\right. .
\end{eqnarray}
The shear is chosen as, 
\begin{eqnarray}
 &  & \sigma_{ij}=\mbox{diag}\left[-\frac{2\sigma_{+}}{H},\frac{\sigma_{+}+\sqrt{3}\sigma_{-}}{H},\frac{\sigma_{+}-\sqrt{3}\sigma_{-}}{H}\right],\label{sigma}
\end{eqnarray}
in this present work.

Metricity together with the condition for zero torsion imply that the spatial part of the connection can be written as 
\begin{equation}
\Gamma_{ijk}=\frac{1}{2}\left\{ \gamma_{ijk}-\gamma_{jik}-\gamma_{kij}\right\}, \label{conexao}
\end{equation}
where $\gamma_{ij}^{k}e_{k}=[e_{j},e_{i}]$ is the commutator of the appropriate base vectors for the group of motions under consideration. The possible groups of motion are classified according to the commutator, which give rise to all Bianchi types \cite{wainwright2005dynamical}. 

In this present work, only strict zero spatial curvature is going to be addressed, so that all commutators are zero, which imply that all the purely spatial components of the connection are null. 

As usual, the Riemann tensor follows from the commutator of the covariant derivatives 
\begin{eqnarray*}
 &  & [\nabla_{c},\nabla_{d}]V^{a}=R_{\, bcd}^{a}V^{b}\\
 &  & R_{bcd}^{a}=\Gamma_{\, bd|c}^{a}-\Gamma_{\, bc|d}^{a}+\Gamma_{\, ec}^{a}\Gamma_{\, bd}^{e}-\Gamma_{\, ed}^{a}\Gamma_{\, bc}^{e}+\gamma_{\, cd}^{e}\Gamma_{\, be}^{a}.
\end{eqnarray*}
In this case the Jacobi identify is satisfied identically. 

This is the anisotropic generalization of the flat Friedmann model, for zero $3-$ curvature $^{3}R=0$.

\section{Field equations\label{s2}}

Metric variation on the theory in \eqref{acao} gives the fields equations
\begin{eqnarray}
&&E_{ab}\equiv \frac{1}{G}\left(G_{ab}+\frac{1}{2}g_{ab}\Lambda\right)\nonumber\\
&&+\left(\beta-\frac{1}{3}\alpha\right)H_{\: ab}^{(1)}+\alpha H_{\: ab}^{(2)}-8\pi T_{ab}=0,\label{eq.campo}
\end{eqnarray}
where 
\begin{eqnarray*}
 &  & G_{ab}=R_{ab}-\frac{1}{2}g_{ab}R,\\
 &  & H_{ab}^{(1)}=\frac{1}{2}g_{ab}R^{2}-2RR_{ab}-2g_{ab}\square R+2R_{;ab},\\
 &  & H_{ab}^{(2)}=\frac{1}{2}g_{ab}R^{cd}R_{cd}-\square R_{ab}-\frac{1}{2}g_{ab}\square R+R_{;ab}\\
 &  & -2R^{cd}R_{cbda}.
\end{eqnarray*}
Let us emphasize that every Einstein space satisfying $R_{ab}=g_{ab}\Lambda/2$, with $T_{ab}=0$,
is an exact solution of \eqref{eq.campo}. All vacuum solutions of
Einstein's field equations are also exact solutions of \eqref{acao}. In this present work, the cosmological constant is set to zero. 

Also, as a metric theory, the covariant divergence $\nabla^{c}E_{ca}=0$.
Any source that satisfies $\nabla^{c}T_{ca}=0$ can be consistently
added in \eqref{eq.campo}. The $00$, $E_{00}=0$ equation is a constraint 
and is given in the Appendix \ref{appendix}, while the $0i$ constraints $E_{0i}=0$ are identically satisfied. These constraints are all dynamically preserved, and the $E_{00}$ constraint is used to numerically check our result. 
\subsection{Einstein gravity \label{EG}}
We will define the expansion normalized variables (ENV) \cite{wainwright2005dynamical},
first in the context of Einstein's GR for which $\alpha=0$ and $\beta=0$
in the field equations \eqref{eq.campo}. The shear $\sigma_{\pm}$ and energy density $T_{00}=\rho$
given respectively in \eqref{conexao_0} and \eqref{eq.campo}, are both divided by appropriate powers of $H$ giving
the new variables $\Sigma_{\pm}$, and $\Omega_{m}$ 
\begin{eqnarray}
 &  & \Sigma_{\pm}=\frac{\sigma_{\pm}}{H}\nonumber \\
 &  & \Omega_{m}=\frac{8\pi G\rho}{3H^{2}}.\label{Omegas}
\end{eqnarray}
In accordance to the tetrad chosen in \eqref{metrica} and the set of variables, the source is 
\begin{equation}
8\pi GT_{ab}=\mbox{diag} [3\Omega_m,3wH^2\Omega_m,3wH^2\Omega_m,3wH^2\Omega_m], \label{tem}
\end{equation}
the vanishing of the covariant divergence of \eqref{tem}, results in the time evolution of $\Omega_m$
\begin{equation}
\dot{\Omega}_m=-2\frac{\dot{H}}{H}\Omega_m-3(w+1)\Omega_m.\label{e.t.Omegas}
\end{equation}
The rest of the ENV are zero since we are
restricting to the Bianchi I case. 

For $\alpha=0$ and $\beta=0$ the field equations \eqref{eq.campo}
reduce to Einstein's equations 
\footnote{these equations coincide with \cite{wainwright2005dynamical} pgs. 114 and 135%
} with matter source
\begin{eqnarray}
 &  & \dot{H}=-(1+q)H\nonumber \\
  &  & \dot{\Sigma}_{+}=-3\Sigma_{+}+(1+q)\Sigma_{+}\nonumber \\
 &  & \dot{\Sigma}_{-}=-3\Sigma_{-}+(1+q)\Sigma_{-}\nonumber \\
 &  & 1=\Sigma^{2}+\Omega_{m},\label{eq.Einstein}
\end{eqnarray}
where $\Sigma^{2}=\sigma_{ij}\sigma^{ij}/(6H^{2})=\Sigma_{+}^{2}+\Sigma_{-}^{2}$,
and 
\[
q=\frac{1}{2}\left(1+3\Sigma^{2}+3w\Omega_{m} \right).
\]
For strict Einstein gravity, \eqref{e.t.Omegas} is given by
\footnote{this equation coincides with \cite{wainwright2005dynamical} pg. 115} 
\[
\dot{\Omega}_{m}=(2+2q-3(w+1))\Omega_{m}, 
\]
which upon substitution of the above expression for $q$ and \eqref{eq.Einstein} results that the time evolution of the density of the Universe is totally independent of the geometry, at least for Bianchi I case
\[
\dot{\Omega}_{m}=3[1-w+(w-1)\Omega_m]\Omega_m.
\]
 We stress that this is a particular feature of pure Einstein gravity, and will not be verified in quadratic gravity for instance, as we shall see in the following. Also concerning Einstein gravity, specifically, the first equation of \eqref{eq.Einstein} can be written as
\begin{equation}
\dot{H}=-\frac{3}{2}\left(w+1+(1-w)\Sigma^2)\right)H. \label{ind.omega1}
\end{equation}
This equation together with 
\begin{eqnarray}
  &  & \dot{\Sigma}_{+}=\frac{3(w-1)}{2}\left(1-\Sigma^2\right)\Sigma_{+}\nonumber \\
 &  & \dot{\Sigma}_{-}=\frac{3(w-1)}{2}\left(1-\Sigma^2\right)\Sigma_{-} 
 \label{ind.omega2},
\end{eqnarray}
define a dynamical system. First, according to \eqref{ind.omega1} and \eqref{ind.omega2} there is a dependence on the parameter $w$ in the sense that the geometry of the Universe depends on the particular equation of state of the substance this Universe is made of. So according to GR, the geometry of the Universe will depend on it's matter content even very close to the physical singularity $H\rightarrow \infty$. This property  will not be satisfied for quadratic gravity, for example, as we shall see in section \ref{s3}. Also, although \eqref{ind.omega1} and \eqref{ind.omega2} do not depend on $\Omega_m$ explicitly, the initial value of $\Sigma^2=\Sigma_{+}^{2}+\Sigma_{-}^{2}$ does depend implicitly on $\Omega_m$ through the constraint equation $1=\Sigma^{2}+\Omega_{m}$. 

\section {Past attractors in quadratic gravity\label{s3}}
Now we turn to the more general case for which $\alpha\neq 0$ and
$\beta\neq0.$ Besides the coordinates defined in \eqref{Omegas}, there
will be higher derivatives, so that, following Barrow and Hervik \cite{Barrow:2006xb}, these
additional ENV are needed 
\begin{eqnarray}
 &  & \Sigma_{\pm1}=\frac{\dot{\sigma}_{\pm}}{H}\nonumber\\
 &  & \Sigma_{\pm2}=\frac{\ddot{\sigma}_{\pm}}{H}\nonumber\\
 &  & Q_{1}=\frac{\dot{H}}{H}\nonumber\\
 &  & Q_{2}=\frac{\ddot{H}}{H}\nonumber\\
 &  & B=\frac{1}{3\beta H^{2}}.\label{Sigma}
\end{eqnarray}
These variables are not strictly identical to that chosen by Barrow and Hervik, since they use proper time instead of dynamical time, which is our choice in this present work.
According to their own definition, the following differential equations
must be satisfied 
\begin{eqnarray}
 &  & \dot{\Sigma}_{\pm}=\Sigma_{\pm1}-\Sigma_{\pm}Q_{1}\nonumber\\
 &  & \dot{\Sigma}_{\pm1}=\Sigma_{\pm2}-\Sigma_{\pm1}Q_{1}\nonumber\\
 &  & \dot{B}=-2Q_{1}B\nonumber\\
 &  & \dot{Q}_{1}=Q_{2}-Q_{1}^{2}.\label{e.t.Sigma}
\end{eqnarray}
The dynamical system is defined be equations \eqref{e.t.Omegas}, \eqref{e.t.Sigma} and the differential equations shown in the Appendix \ref{appendix}. We have redefined the constant $\alpha$ in \eqref{acao}  as $\alpha=3\beta\chi$. 
Remind the curvature scalar 
\begin{eqnarray*}
&&R_{ab}R^{ab}=\left\{4\Sigma_{{-1}}\Sigma_{{-}}+4\left(\Sigma_{{-}}\right)^{4}+4+4\left(\Sigma_{{+}}\right)^{4}\right.\\
&&\left.+4Q_{{1}}\left(\Sigma_{{-}}\right)^{2}+10\left(\Sigma_{{-}}\right)^{2}+10\left(\Sigma_{{+}}\right)^{2}+4\Sigma_{{+1}}\Sigma_{{+}}\right.\\
&&\left.+4Q_{{1}}+\frac{2}{3}\left(\Sigma_{{-1}}\right)^{2}+8\left(\Sigma_{{-}}\right)^{2}\left(\Sigma_{{+}}\right)^{2}+4Q_{{1}}\left(\Sigma_{{+}}\right)^{2}\right.\\
&&\left.+\frac{2}{3}\left(\Sigma_{{+1}}\right)^{2}+\frac{4}{3}Q_{{1}}^{2}\right\}/(\beta B)^2, 
\end{eqnarray*}
so that $B\rightarrow 0$ is a physical singularity. 
The advantage of using expansion normalized variables is that they remain finite through the singularity $B\rightarrow 0$. Albeit curvature invariants diverge, they do not appear explicitly in the dynamical system. 

In this section we describe possible past attractors. It is useful to note that  by inspection of the differential equations in Appendix \ref{appendix} it can be seen the presence of  $\Omega_m$ only in the equation for $\dot{Q}_2$ through the product $\Omega_m B$. So that this product, could be irrelevant when  $B\rightarrow 0$. In this regime the matter content decouples from the evolution of the Universe, 
in a strong sense. For a pure quadratic gravity  when $B=0$, the geometry of the Universe is absolutely independent of the matter content, or EOS of which this Universe is made of. This is a drastic difference between Einstein gravity described in section \ref{EG} and pure quadratic gravity. 

While the geometry of the Universe does not depend on it's matter content, the time evolution of the density parameter given by \eqref{e.t.Omegas}, does depend on the geometry through $\dot{H}/H=Q_1$, as will become clear in section \ref{A}.
\begin{figure}[htpb]
 \begin{center}
  \resizebox{\imsize}{!}{\includegraphics{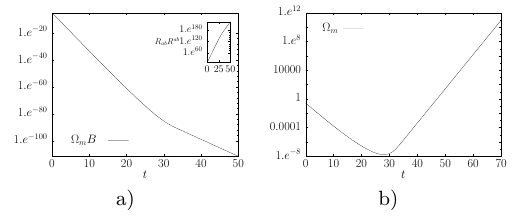}}  
    \end{center}
    \caption{In this orbit we choose $\chi=3$, $\beta=1.2$, $\Omega_m=0.23$, $\phi=2.43$ and $w=0.7$, the complete specification of the initial condition is given in Appendix \ref{appendix1}. The evolution of the system is to the past and  a) shows that the product $\Omega_mB$ vanishes as the singularity is approached while b) shows that $\Omega_m$ diverges. In the inset of a) it is shown the product of the curvature invariant $R_{ab}R^{ab}$. \label{fig9}}
\end{figure}

This means that all the
results of Barrow and Hervik  \cite{Barrow:2006xb} considering vacuum pure quadratic regime ($B=0$) hold also in the presence of matter and we can the use results
from this paper.
\subsection{Isotropic singularity \label{A}}
This is an exact solution of QG which was found by Barrow and Hervik \cite{Barrow:2006xb} and corresponds in the limit $t\rightarrow -\infty$ to the fixed point with the only non null variables 
\begin{eqnarray*}
&&Q_1=-2\\
&&Q_2=4.
\end{eqnarray*}
It is a vacuum solution $T_{ab}=0$, that behaves as if it was a strict GR universe filled with radiation $p=\rho /3$. The eigenvalues together with the degeneracies are 
\[
2_2,\, 1-3w,\, 5,\, 4,\, 3_3, 1_2
\]
So that it is an attractor to the past for $w<1/3$. In the opposite case this vacuum solution is a saddle. Our numerical studies show 
that
for the EOS parameter $w<1/3$, $\Omega_m$ decreases towards the singularity, while for $w>1/3$, $\Omega_m$ increases and finally diverges. FIG. \ref{fig9}, shows a solution for which $\Omega_m$ increases, although the product $\Omega_m B\to 0$. The latter fact explains why all other variables
have the same asymptotic as for vacuum isotropic solution, since matter density enters equation of motion only with this product. As these two asymptotics differs only in the behavior of $\Omega_m$ we will refer to both of them as an isotropic solution in what follows. Our numerics
indicates also that for the particular case of $w=1/3$ the variable  $\Omega_m$ tends to some constant finite value.
 
\subsection{Kasner solution \label{B}}
This is also an exact solution of QG and in the ENV coordinates it is is given by  $\Omega_{\Lambda}=0,$
\begin{eqnarray*}
 &  & B=\frac{3e^{6t}}{\beta}\\
 &  & \Sigma_{+}=\cos\phi\\
 &  & \Sigma_{-}=\sin\phi,
\end{eqnarray*}
where $\phi$ is a constant parameter. In this case the asymptotic
$t\rightarrow -\infty$ it is also a fixed point whose eigenvalues and degeneracies are 
\[
\,9,\,3-3w,\,0_{2},\,6_{6}.
\]
This solution is also an attractor to the past except for these two zero eigenvalues. One zero eigenvalue appears because we have
in fact an one-dimensional set of equilibrium points (the Kasner circle) marked by the parameter $\phi$. The second zero eigenvalue
indicates that the fixed points of this set are non-hyperbolic fixed points and their stability requires a special analysis. We will see
that in some situations they can be attractors.

\section{Initial conditions\label{s4}}
Only backwards evolution is analyzed, in the sense that the system evolves to a singularity, and as is known, there are two solutions which are attractors to the past. As for initial conditions, since the phase space of the system under investigation is rather high dimensional,
we need some principle to fix reasonable starting points of our numerical integrations instead of trying to cover all possible initial
conditions. That is why our 
 initial condition is chosen near a GR solution, in a sense that we will make clear in the following. We stress that the initial value for the density parameter $\Omega_m$ is going to be related to the amount of anisotropy, the closer $\Omega_m$ is to $1$ the smaller the amount of anisotropy. 

The general GR solution for Bianchi I and dust was discovered independently by Robinson \cite{robinson1961relativistic} and Heckmann Shc\"ucking \cite{heckmann1962gravitation} and generalized to arbitrary equation of state by Jacobs \cite{jacobs1968spatially}.
Strictly speaking, this is an exact solution for GR, with equation of state $p=w \rho$ 
\begin{eqnarray}
&&H=\frac{\left( \alpha+{m}^{2}{\tau}^{1-w} \right) ^{{\frac {w}{-1+w}}} \left( 
\alpha\,{\tau}^{w}+2\,{m}^{2}\tau \right) 
}{3\,\tau \left( \alpha\,{\tau}^{w}+{m}^{2}\tau \right)}\nonumber\\
&&\Omega_m=4\,{\frac {{m}^{2}\tau \left( \alpha\,{\tau}^{w}+{m}^{2}\tau \right) }{ \left( 
\alpha\,{\tau}^{w}+2\,{m}^{2}\tau \right) ^{2}}}\nonumber\\
&&\Sigma_+=-\,{\frac {{\tau}^{-1+w}\alpha\cos(\phi) \tau}{\alpha
\,{\tau}^{w}+2\,{m}^{2}\tau}}\nonumber\\
&&\Sigma_-=-\,{\frac {{\tau}^{-1+w}\alpha\sin(\phi)\tau}{
\alpha\,{\tau}^{w}+2\,{m}^{2}\tau}} \label{jacobs}
\end{eqnarray}
where $\tau$ is related to the dynamical time $t$ by, 
\[
t=-\,{\frac {-2\,w\ln  \left( \tau \right) +\ln  \left( \tau \right) +\ln  \left( \alpha\,{\tau}^{w}+{m}^{2}\tau \right) }{3(-1+w),
}}
\]
the rest of the variables and the specification of the initial condition are presented in the Appendices. 

As mentioned above, this solution is an exact solution of GR with non zero source. Unless the EOS parameter $w=-1$, it will not be a solution of quadratic gravity. However,  for sufficiently big values of $B$ the initial condition is not so close to the singularity, the Universe evolves according to GR and the quadratic terms behave as a perturbation, locally in time. That is why we have chosen our initial conditions from which we integrate
numerically our system to be near Jacobs solution. Namely, we choose all variable satisfying Jacobs solution except for $Q_2$ which is found from the constraint equation, $E_{00}$, shown Appendix \ref{appendix}. The initial condition diferes from GR only in higher time derivatives.

Remind that $B\rightarrow\infty $ is a coordinate singularity, and in order to correctly understand this regime, a different coordinate system should be chosen.

Stiff matter, with equation of state $w=1$, is also a GR exact solution which is also a fixed point 
\begin{eqnarray}
&&\Sigma_+=-{\frac {\alpha\,\cos \left( \phi \right) }{\alpha+2\,{m}^{2}}}\nonumber\\
&&\Sigma_-=-{\frac {\alpha\,\cos \left( \phi \right) }{\alpha+2\,{m}^{2}}}\nonumber\\
&&\Omega_m=4\,{\frac {{m}^{2} \left( \alpha+{m}^{2} \right) }{ \left( \alpha+2\,{
m}^{2} \right) ^{2}}}\nonumber\\
&&Q_1=-3\nonumber\\
&&Q_2=9\\
&&\Sigma_{+1}=-3\Sigma_+\nonumber\\
&&\Sigma_{-1}=-3\Sigma_-\nonumber\\
&&\Sigma_{+2}=9\Sigma_+\nonumber\\
&&\Sigma_{-2}=9\Sigma_-, \label{stiff.matter}
\end{eqnarray}
again $Q_2$ is going to be fixed by the constraint $E_{00}$. The relation between the dynamical time $t$ and $\tau$ and the $H$ variable are not the same:
\begin{eqnarray}
&&H=\frac{(1+2m)}{3\tau^{1+2m}}\nonumber\\
&&t=\frac{(1+2m)\ln (\tau)}{3}\label{stiff.matter.H}
\end{eqnarray}

\section{Numerical results\label{s5}}

\subsection{Stiff fluid}

\begin{figure}[htpb]
 \begin{center}
  \resizebox{\imsize}{!}{\includegraphics{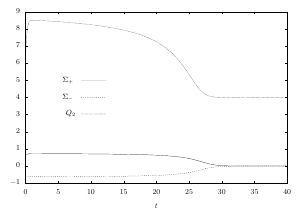}}  
    \end{center}
    \caption{In this orbit we choose $\chi=3$, $\beta=1.2$, $\Omega_m=0.23$, $\phi=2.43$ and $w=1$, the complete specification of the initial condition is given in Appendix \ref{appendix1}, and in accordance to \eqref{stiff.matter.H}. The evolution of the system is to the past and the solution approaches the isotropic singularity to the past. \label{fig7}}
\end{figure}

In the case of stiff fluid the ENV of Jacobs solution are constants, so we can formally start to integrate in the pure quadratic regime
with $B=0$. This allows us to use second order stability analysis of Kasner solution made in \cite{Barrow:2006xb}. Indeed, following this
paper we can see that
the Kasner circle should be a saddle node fixed point with $X\rightarrow 0$
\[
\dot{X}=X^2+\mathcal{O}(X^3)
\]
where 
\begin{equation}
X=\left( \frac{14+\chi}{72}\right)x-\left( \frac{\chi-4}{6}\right)y+\left( \frac{5\chi -2}{72}\right)z+\left(\frac{2+\chi}{72} \right)\tilde{w}, \label{center.manifold}
\end{equation}
where 
\begin{eqnarray*} 
&& (\Sigma_+,\Sigma_-)=\Sigma (\cos \phi,\sin\phi)\;\;\;\;(\Sigma_{+1},\Sigma_{-1})=F\Sigma(\cos \phi,\sin\phi) \\
&&(\Sigma_{+2},\Sigma_{-2})=G\Sigma(\cos \phi,\sin\phi)\\
&&(Q_1,\Sigma,F,G)=(-3+x,-1+y,3+z,-9+\tilde{w}).
\end{eqnarray*}
Exact Kasner solution corresponds to $x=y=z=\tilde{w}=0$. Now we consider the initial condition specified in Appendix \ref{appendix1}. The limit in dynamical time $t=-\infty$, corresponds to $\tau=0$ for the GR solution shown in Appendix \ref{appendix1}, when the EOS parameter $w\rightarrow 1$
\begin{eqnarray*} 
&&x=0\\
&&y=\frac{2m^2}{\alpha}\\
&&z=\frac{-6m^2}{\alpha}\\
&&w=\frac{18m^2}{\alpha}.
\end{eqnarray*}
By direct substitution of these expressions into \eqref{center.manifold} results in 
\[
X=-\left( \frac{\chi-4}{6}\right)(2)+\left( \frac{5\chi -2}{72}\right)(-6)+\left(\frac{2+\chi}{72} \right)(18),
\]
which is zero, $X=0$, for 
\[
\chi=4.
\] 
The initial condition is specified in the Appendix \ref{appendix1}, while the value of $B=1/(3\beta H^2)$ is set up by equation \eqref{stiff.matter.H} for matching with the next subsection. We have checked that the results are practically indistinguishable
from those with initial $B=0$, so we do not present them separately. In FIG. \ref{fig7}, we present a solution with the initial condition for $\Omega_m=0.23$. and coupling constant $\chi=3$. It can be seen that this solution shown in FIG. \ref{fig7} approaches the isotropic singularity to the past. 

\begin{figure}[htpb]
 \begin{center}
  \resizebox{\imsize}{!}{\includegraphics{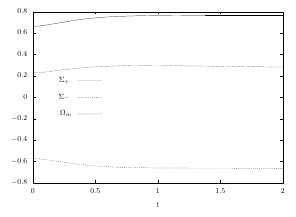}}  
    \end{center}
    \caption{In this orbit we choose $\chi=6$, $\beta=1.2$, $\Omega_m=0.23$, $\phi=2.43$ and $w=1$, the complete specification of the initial condition is given in Appendix \ref{appendix1}, and in accordance to \eqref{stiff.matter.H}. The evolution of the system is to the past and the solution approaches the Kasner singularity to the past.\label{fig8}}
\end{figure}

\begin{figure}[htpb]
\begin{center}
   \resizebox{\imsize}{!}{\includegraphics{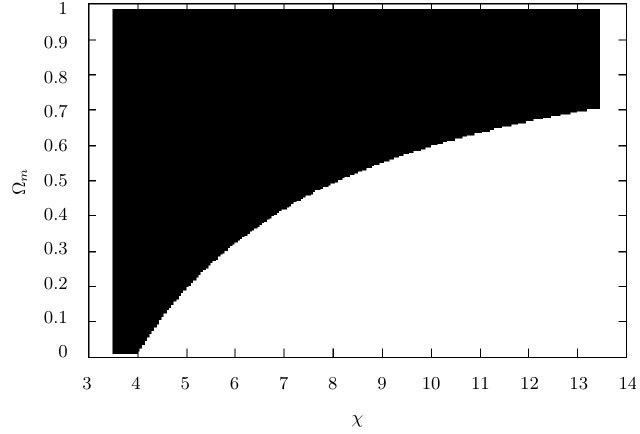}}  
    \end{center}
    \caption{A grid of $100\times 100$ for values of $\Omega_m$ and $\chi$ with $\beta=1.2$, $w=1$, $\phi=2.43$. The evolution of the system is to the past and every black point corresponds to a solution that approaches the isotropic singularity while the white points approach Kasner solution. It can be seen that there's a bifurcation at $\chi\simeq 4$, for $\Omega_m\simeq 0$ which corresponds to an initial condition near Kasner solution.\label{fig4}}
\end{figure}

In FIG. \ref{fig8} we choose a different value for the coupling constant, $\chi=6$ while keeping all the other numerical values. In this case, it can be seen that the solution shown in FIG. \ref{fig8}, approaches the Kasner solution to the past.
In FIG. \ref{fig4} we show a grid of initial conditions, considering a Universe initially filled with stiff fluid. It can be seen that there's a bifurcation at $\chi\simeq 4$, for $\Omega_m\simeq 0$ which corresponds to an initial condition near Kasner solution, showing agreement with the center manifold analysis above. 
\subsection{A general perfect fluid} 
\begin{figure}[htpb]
 \begin{center}
  \resizebox{\imsize}{!}{\includegraphics{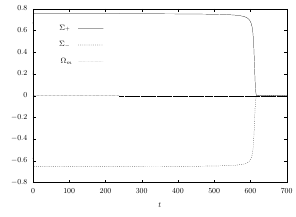}}  
    \end{center}
    \caption{In this orbit we choose $\chi=3$, $\beta=1.2$, $\Omega_m=0.23$, $\phi=2.43$ and $w=-0.9.$ The evolution of the system is to the past and the solution approaches the isotropic singularity, just as FIG. \ref{fig7}.\label{fig2}}
\end{figure}

\begin{figure}[htpb]
\begin{center}
   \resizebox{\imsize}{!}{\includegraphics{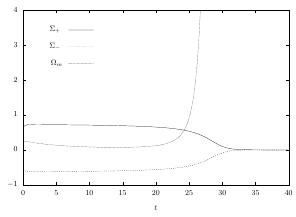}}  
    \end{center}
    \caption{In this orbit we choose $\chi=3$, $\beta=1.2$, $\Omega_m=0.23$, $\phi=2.43$ and $w=0.9.$ The evolution of the system is to the past and the solution approaches the isotropic singularity.\label{fig3}}
\end{figure}

The intention of this subsection is to show that the result is not much sensitive to the type of fluid which we initially choose to fill the Universe. In FIG. \ref{fig2} an initial condition with a specific value of the coupling constant $\chi=3$, while in FIG. \ref{fig3} the value of $\chi=0.9$. Both FIGS. \ref{fig2} and \ref{fig3} show that this universe has an isotropic beginning after which anisotropies grow, and a solution near Jacobs is obtained today. 
It can be seen that in FIGS. \ref{fig2}, \ref{fig3} and \ref{fig7} the asymptotic past of the Universe is the isotropic singularity, so that the EOS parameter does not influences much the behavior of the Universe near the singularity.   

\begin{figure}[htpb]
\begin{center}
   \resizebox{\imsize}{!}{\includegraphics{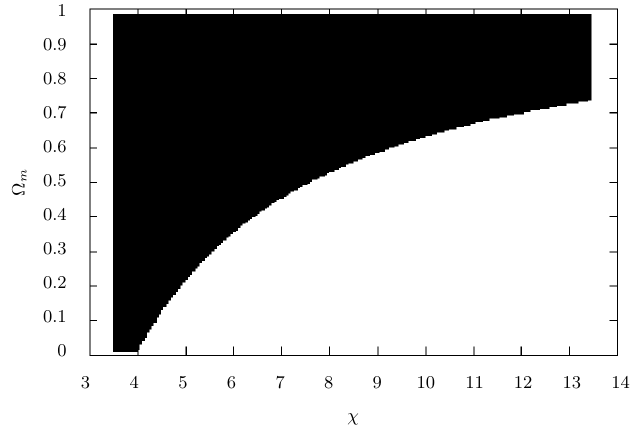}}  
    \end{center}
    \caption{A grid of $100\time 100$ values $\Omega_m \times \chi$ with $\beta=1.2$, $w=0.5$, $\phi=2.43$. The evolution of the system is to the past and again every black point corresponds to a solution that approaches the isotropic singularity while the white points approach Kasner solution. By comparing with FIG. \ref{fig4}, it can be seen that the asymptotic dynamical structure of the singularity is not too much influenced by the initial substance the Universe.\label{fig5}}
\end{figure}

In FIG. \ref{fig5} we show a bifurcation diagram for an initial Universe near Jacobs solution with an EOS parameter $w=0.5$. 
Similar diagrams have been constructed for other $w$ in the range $-1<w<1$, they are all qualitatively the same.
The value of the ENV $B$ is initially set to be very small, so FIG. \ref{fig5} shows the behavior of the universe very deep into the singularity.
We have checked also that the results do not change much for much bigger initial $B$ except for the case of extremly low matter
density where some small changes have been detected. More work will be needed to understand whether this is a real efect of just
a numerical instability. 

  Comparing to FIGS. \ref{fig4} and \ref{fig5}, it can be seen that the initial substance of which the Universe is made does not influences the bifurcation value of $\chi$.
\\

\newpage
\section{Conclusions}
In the present paper we considered cosmological dynamics towards a singularity in quadratic gravity for a Bianchi I anisotropic
Universe. Two important features distinguishing quadratic gravity case from the GR analog are:
\begin{itemize}

\item A new stable isotropic regime
appears.

\item Anisotropic Kasner solution looses a linear stability becoming a saddle-node
fixed point.
\end{itemize}

 For initial conditions which we use in the present paper - those in the vicinity of Jacobs GR solution - the dynamics 
near Kasner solution appears to be determined by the value of coupling constant $\chi$. Note, that assuming that evolution
of the Universe during preceding GR era (on which quadratic terms can be considered as tiny perturbations) is governed by Jacobs
solution results into a Universe which is very close to Kasner solution at the beginning of a quadratic gravity epoch. For such initial conditions
and $\chi < 4$ the Kasner circle appears to be unstable. Our numerical analysis shows that independently of the relative amount of 
matter at the start of quadratic gravity era, the solution tends to the isotropic attractor. This also makes a chaotic attractor
like BKL oscillations for more general anisotropic models highly improbable in quadratic gravity, though this issue requires further
investigations as we are restricted by Bianchi I model in the present paper. Nevertheless, instability of Kasner solution can be currently
considered as a hint for instability of BKL oscillations in quadratic gravity.

On the contrary, for $\chi>4 $ for Jacobs initial conditions the Kasner circle is locally stable. Numerical integrations show that for 
such $\chi$ there are two types of dynamics. If matter content at a period when quadratic contribution becomes important is low enough,
the Kasner circle is still a future attractor. For bigger $\Omega_m$ the solution tends to the isotropic attractor. This value of matter
density which switches the nature of singularity from anisotropic to isotropic one is  a growing function of coupling constant $\chi$.
We can also see that the boundary between these two regimes in the initial condition space is sharp, so apparently the dynamics has no chaotic
properties. 

Note also that the case of stiff fluid in quadratic gravity is not exceptional in contrast to GR. In GR the relative matter density of a stiff
fluid does not disappear when the Universe evolves towards singularity while $\Omega_m \to 0$ for $w<1$. In quadratic gravity
the dynamics for $w=1$ has no qualitative difference from the $-1<w<1$ case.

\begin{acknowledgements} 
D. M. would like to thank CAPES grant 8772-13-4, and the kind hospitality at CWRU were part of this work was done, and the Brazilian project ``Nova F\'\i sica no Espa\c co". The work of A.T.was supported by RFBR Grant 14-02-00894 and partially supported by the Russian 
Government Program of Competitive Growth of Kazan Federal University. A.T. thanks Instituto de Fisica, Universidade de Brasilia,
where part of this work was done, for hospitality.
\end{acknowledgements} 
\newpage
\appendix

\section{The initial condition for Jacobs solution \label{appendix1}}
\begin{widetext}
\begin{eqnarray}
&  &Q_1= \left[-3\,{\alpha}^{4}{\tau}^{4\,w}-12\,{\tau}^{3\,w+1}{\alpha}^{3}{m}^{2}-21\,{\tau}^{2+2\,w}{\alpha}^{2}{m}^{4}-18\,{\tau}^{w+3}\alpha\,{m}^{6}-6\,w{m}^{2}{\tau}^{3\,w+1}{\alpha}^{3}\right.\nonumber\\
 &  & \left.-18\,w{m}^{4}{\tau}^{2+2\,w}{\alpha}^{2}-18\,w{m}^{6}{\tau}^{w+3}\alpha-6\,{\tau}^{4}{m}^{8}-6\,{\tau}^{4}w{m}^{8}\right]\left[ \left( \alpha\,{\tau}^{w}+2\,{m}^{2}\tau \right) ^{2} \left( \alpha\,{\tau}^{w
}+{m}^{2}\tau \right) ^{2}
\right]^{-1}\nonumber\\
&&Q_2= 9\,{\tau}^{2}\left[24\,{w}^{2}{m}^{14}{\tau}^{w+5}\alpha+120\,w{m}^{8}{\tau}^{2+4\,w}{\alpha}^{4}+120\,w{m}^{12}{\tau}^{4+2\,w}{\alpha}^{2}+160\,w{m}^{10}{\tau}^{3+3\,w}{\alpha}^{3}+\right.\nonumber\\
 &  & 48\,w{m}^{14}{\tau}^{w+5}\alpha+48\,w{m}^{6}{\tau}^{1+5\,w}{\alpha}^{5}+90\,{\tau}^{3+3\,w}{\alpha}^{3}{m}^{10}{w}^{2}+4\,{m}^{16}{\tau}^{6}+24\,{\tau}^{w+5}\alpha\,{m}^{14}+10\,{\tau}^{7\,w-1}{\alpha}^{7}{m}^{2}+\nonumber\\
 &  & 8\,w{m}^{16}{\tau}^{6}+4\,{w}^{2}{m}^{16}{\tau}^{6}+66\,{\tau}^{4+2\,w}{\alpha}^{2}{m}^{12}+40\,{\tau}^{6\,w}{\alpha}^{6}{m}^{4}+121\,{\tau}^{2+4\,w}{\alpha}^{4}{m}^{8}+110\,{\tau}^{3+3\,w}{\alpha}^{3}{m}^{10}+\nonumber\\
 &  & 88\,{\tau}^{1+5\,w}{\alpha}^{5}{m}^{6}+{\tau}^{8\,w-2}{\alpha}^{8}+62\,{\tau}^{4+2\,w}{w}^{2}{\alpha}^{2}{m}^{12}+8\,w{m}^{4}{\tau}^{6\,w}{\alpha}^{6}+14\,{\tau}^{6\,w}{\alpha}^{6}{m}^{4}{w}^{2}+\nonumber\\
 &  & \left.80\,{\tau}^{2+4\,w}{\alpha}^{4}{m}^{8}{w}^{2}+2\,{\tau}^{7\,w-1}{\alpha}^{7}{m}^{2}{w}^{2}+44\,{\tau}^{1+5\,w}{\alpha}^{5}{m}^{6}{w}^{2}\right]\left[\left(\alpha\,{\tau}^{w}+2\,{m}^{2}\tau\right)^{4}\left(\alpha\,{\tau}^{w}+{m}^{2}\tau\right)^{4}\right]^{-1}\nonumber\\
 &&\Sigma_{+1}= \frac{3\tau\alpha\,\cos\left(\phi\right)\left({\tau}^{3\,w-1}{\alpha}^{2}+3\,{\tau}^{2\,w}\alpha\,{m}^{2}+2\,{\tau}^{1+w}{m}^{4}\right)}{\left(\alpha\,{\tau}^{w}+{m}^{2}\tau\right)\left(\alpha\,{\tau}^{w}+2\,{m}^{2}\tau\right)^{2}}\nonumber\\
 &&\Sigma_{-1}=\frac{3\tau\alpha\,\sin\left(\phi\right)\left({\tau}^{3\,w-1}{\alpha}^{2}+3\,{\tau}^{2\,w}\alpha\,{m}^{2}+2\,{\tau}^{1+w}{m}^{4}\right)}{\left(\alpha\,{\tau}^{w}+{m}^{2}\tau\right)\left(\alpha\,{\tau}^{w}+2\,{m}^{2}\tau\right)^{2}}\nonumber\\
 &&\Sigma_{+2}=-9\,{\frac{\tau\alpha\,\cos\left(\phi\right)\left(8\,{\tau}^{5\,w}{\alpha}^{4}{m}^{2}+8\,{\tau}^{4+w}{m}^{10}+{\tau}^{-1+6\,w}{\alpha}^{5}+38\,{\tau}^{2+3\,w}{\alpha}^{2}{m}^{6}+28\,{\tau}^{3+2\,w}{m}^{8}\alpha+25\,{\tau}^{4\,w+1}{\alpha}^{3}{m}^{4}\right)}{\left(\alpha\,{\tau}^{w}+{m}^{2}\tau\right)^{2}\left(\alpha\,{\tau}^{w}+2\,{m}^{2}\tau\right)^{4}}}\nonumber\\
 &&\Sigma_{-2}=-9\,{\frac{\tau\alpha\,\sin\left(\phi\right)\left(8\,{\tau}^{5\,w}{\alpha}^{4}{m}^{2}+8\,{\tau}^{4+w}{m}^{10}+{\tau}^{-1+6\,w}{\alpha}^{5}+38\,{\tau}^{2+3\,w}{\alpha}^{2}{m}^{6}+28\,{\tau}^{3+2\,w}{m}^{8}\alpha+25\,{\tau}^{4\,w+1}{\alpha}^{3}{m}^{4}\right)}{\left(\alpha\,{\tau}^{w}+{m}^{2}\tau\right)^{2}\left(\alpha\,{\tau}^{w}+2\,{m}^{2}\tau\right)^{4}}}\nonumber\\
 &&\label{eq.gr}
\end{eqnarray}
\end{widetext}

The initial condition is chosen as follows. First, we fix a numerical value for the time $\tau=1\times 10^{-6}.$ The closer $\tau$ is to zero, the closer to the curvature singularity. Then, as the parameter $m$ can be chosen arbitrarily, we set it as $m=1$. The value of $\Omega_m$ is used to set up the value of the parameter $\alpha$ as
\[
\alpha=-2\,{\frac {{m}^{2}\tau}{{\tau}^{w}}}+2\,{\frac {{m}^{2}\tau}{\Omega_{{m}}{\tau}^{w}}}+
2\,{\frac {{m}^{2}\sqrt {-\Omega_{{m}}{\tau}^{2} \left( {\tau}^{w} \right) ^{2}+{
\tau}^{2} \left( {\tau}^{w} \right) ^{2}}}{\Omega_{{m}} \left( {\tau}^{w} \right) ^{
2}}}.
\]
The numerical values of $\alpha$, $m$, $\tau$ and $w$ are substituted back into \eqref{jacobs} and the above equations \eqref{eq.gr}, except for $Q_2$ which is fixed by the constraint $E_{00}$ shown in Appendix \ref{appendix}.
\newpage
\section{The differential equations \label{appendix}}
\begin{widetext}
\begin{eqnarray*}
 &  & \frac{d}{dt}Q_{2}=\left\{ -3\Sigma_{-}^{2}G-\frac{3}{2}\Sigma_{+}^{4}G-\frac{21}{2}Q_{1}^{2}G+\frac{3}{4}\Sigma_{-}^{2}B+\frac{3w}{4}\Omega_{m}B
  +\frac{3}{4}\Sigma_{+}^{2}B+\frac{1}{2}Q_{1}B-\frac{3}{2}\Sigma_{-}^{4}G-2\Sigma_{+1}^{2}G-Q_{1}^{3}G-3\Sigma_{+}^{2}G\right.\\
 &  & -9Q_{1}G-6Q_{2}G-2\Sigma_{-1}^{2}G+\frac{1}{2}\chi Q_{1}\Sigma_{-}\Sigma_{-1}G+\frac{1}{2}\chi\Sigma_{+}Q_{1}\Sigma_{+1}G+\frac{3}{4}B
  -2\Sigma_{-}^{2}Q_{1}G-2\Sigma_{+}^{2}Q_{1}G+\frac{3}{4}\chi\Sigma_{+}^{2}G+\frac{3}{4}\chi\Sigma_{-}^{2}G\\
 &  & -\frac{1}{4}\chi\Sigma_{+1}^{2}G-5Q_{1}Q_{2}G-2\Sigma_{+}\Sigma_{+2}G-4\Sigma_{+}\Sigma_{+1}G
  -4\Sigma_{-}\Sigma_{-1}G-3\chi\Sigma_{-}^{4}G-3\Sigma_{+}^{2}\Sigma_{-}^{2}G-2\Sigma_{-}\Sigma_{-2}G-\frac{1}{4}\chi\Sigma_{-1}^{2}G\\
 &  & +\frac{1}{2}\chi\Sigma_{+}^{2}Q_{1}G+\chi\Sigma_{+}\Sigma_{+1}G-2Q_{1}\Sigma_{+}\Sigma_{+1}G-2Q_{1}\Sigma_{-}\Sigma_{-1}G-3\chi\Sigma_{+}^{4}G-6\chi\Sigma_{+}^{2}\Sigma_{-}^{2}G+\frac{1}{2}\chi\Sigma_{-}^{2}Q_{1}G+\chi\Sigma_{-}\Sigma_{-1}G\\
 &  & \left.+\frac{1}{2}\chi\Sigma_{+}\Sigma_{+2}G+\frac{1}{2}\chi\Sigma_{-}\Sigma_{-2}G\right\} /G\\
 &  & \frac{d}{dt}\Sigma_{-2}=\left\{ -10\chi\Sigma_{-1}Q_{1}G-\chi Q_{2}\Sigma_{-1}G+8\chi\Sigma_{-1}\Sigma_{+}^{2}G-7\chi Q_{1}\Sigma_{-}G-\chi\Sigma_{-}Q_{1}^{2}G-\chi Q_{1}^{2}\Sigma_{-1}G+8\Sigma_{+1}\Sigma_{+}\Sigma_{-}G\right.\\
 &  & -\chi\Sigma_{-}Q_{2}G-4\chi Q_{1}\Sigma_{-2}G+24\chi\Sigma_{-}^{2}\Sigma_{-1}G+24\chi\Sigma_{+}^{2}\Sigma_{-}G+16\chi\Sigma_{+1}\Sigma_{-}\Sigma_{+}G+12\Sigma_{-}^{3}G+8\Sigma_{-1}G-3\Sigma_{-}B-\Sigma_{-1}B\\
 &  &+24\Sigma_{-}G +4\Sigma_{-1}\Sigma_{+}^{2}G+28\, Q_{1}\Sigma_{-}G+24\chi\Sigma_{-}^{3}G+4\Sigma_{-}Q_{2}G-6\chi\Sigma_{-2}G-11\chi\Sigma_{-1}G+12\Sigma_{-}^{2}\Sigma_{-1}G+4\Sigma_{-}Q_{1}^{2}G\\
 & &\left.+12\Sigma_{+}^{2}\Sigma_{-}G-6\chi\Sigma_{-}G+4\Sigma_{-1}Q_{1}G\right\} /(\chi G)\\
 &  & \frac{d}{dt}\Sigma_{+2}=\left\{ -3\Sigma_{+}B-\Sigma_{+1}B-\chi\Sigma_{+}Q_{2}G+12\Sigma_{+}^{3}G+8\Sigma_{+1}G-7\chi Q_{1}\Sigma_{+}G+16\chi\Sigma_{-1}\Sigma_{+}\Sigma_{-}G+8\chi\Sigma_{-}^{2}\Sigma_{+1}G\right. \\
 & &-\chi Q_{2}\Sigma_{+1}G+24\chi\Sigma_{-}^{2}\Sigma_{+}G-\chi Q_{1}^{2}\Sigma_{+1}G+8\Sigma_{-1}\Sigma_{+}\Sigma_{-}G-10\chi\Sigma_{+1}Q_{1}G-4\chi Q_{1}\Sigma_{+2}G+24\chi\Sigma_{+1}\Sigma_{+}^{2}G-\chi\Sigma_{+}Q_{1}^{2}G\\
 &  & +4\Sigma_{+}Q_{2}G-11\chi\Sigma_{+1}G+4\Sigma_{+}Q_{1}^{2}G+28Q_{1}\Sigma_{+}G+4\Sigma_{+1}Q_{1}G+12\Sigma_{-}^{2}\Sigma_{+}G+12\Sigma_{+1}\Sigma_{+}^{2}G+4\Sigma_{-}^{2}\Sigma_{+1}G-6\chi\Sigma_{+2}G\\
 &  &\left. +24\chi\Sigma_{+}^{3}G-6\chi\Sigma_{+}G+24\Sigma_{+}G\right\} /(\chi G)
\end{eqnarray*}
\end{widetext}
subject to the $E_{00}$ constraint 
\begin{widetext}
\begin{eqnarray*}
 &  & \left\{-36GQ_{1}-6GQ_{1}^{2}-12GQ_{2}+36G\Sigma_{+}^{2}+36G\Sigma_{-}^{2}+18G\Sigma_{+}^{4}+18G\Sigma_{-}^{4}-3B\Sigma_{-}^{2}-3B\Sigma_{+}^{2}-3B\Omega_{m}-6G\chi\Sigma_{+}\Sigma_{+2}\right.\\
 &  & -6G\chi Q_{1}\Sigma_{+}^{2}-12G\chi\Sigma_{+1}\Sigma_{+}-6G\chi\Sigma_{-}\Sigma_{-2}-6G\chi\Sigma_{-}^{2}Q_{1}-12G\chi\Sigma_{-1}\Sigma_{-}+72G\chi\Sigma_{-}^{2}\Sigma_{+}^{2}-24G\Sigma_{-1}\Sigma_{-}+36G\Sigma_{-}^{2}\Sigma_{+}^{2}\\
 &  & +3G\chi\Sigma_{+1}^{2}+3G\chi\Sigma_{-1}^{2}-9G\chi\Sigma_{+}^{2}-9G\chi\Sigma_{-}^{2}+36G\chi\Sigma_{+}^{4}+36G\chi\Sigma_{-}^{4}+24GQ_{1}\Sigma_{+}^{2}-24G\Sigma_{+1}\Sigma_{+}+24G\Sigma_{-}^{2}Q_{1}\\
 &  & \left.-6G\chi Q_{1}\Sigma_{+}\Sigma_{+1}-6G\chi Q_{1}\Sigma_{-}\Sigma_{-2}+3B\right\}/B=0
\end{eqnarray*}
\end{widetext}
\bibliographystyle{apsrev4-1}
\bibliography{refs}
\end{document}